\documentclass[10pt,preprint,a4paper]{aastex}

\usepackage{graphics,epsf}
\usepackage{amsmath}                
\usepackage{amsfonts}               
\usepackage{amssymb}                
\usepackage{epsfig}                 
\usepackage{hyperref}
\usepackage{subfigure}

\def \cm{~\rm{cm}}
\def \s{~\rm{s}}
\def \km{~\rm{km}}

\def \K{~\rm{K}}
\def \g{~\rm{g}}

\def \AU{~\rm{AU}}
\def \erg{~\rm{erg}}

\def \yr{~\rm{yr}}

\def \days{~\rm{day}}

\title{IMPULSIVE EJECTION OF GAS IN BIPOLAR PLANETARY NEBULAE}

\author{Muhammad Akashi\altaffilmark{1} and Noam Soker\altaffilmark{1}}

\altaffiltext{1}{Department of Physics, Technion -- Israel Institute of
Technology, Haifa 32000 Israel; akashi@physics.technion.ac.il; soker@physics.technion.ac.il.}

\begin{document}

\begin{abstract}
We simulate the formation of bipolar planetary nebulae (PNe) through very short impulsive mass ejection events from binary systems,
where the asymptotic giant branch (AGB) star ejects a mass shell that is accelerated by jets launched from a compact companion.
The acceleration process takes place at very short distances from the binary system, such that the photon-diffusion time is long enough
to prevent rapid cooling of the shocked jets' material.
When the shocked jets' gas density is lower than the shell density the flow becomes Rayleigh-Taylor unstable and dense clumps are formed in
the flow. At later times a PN with clumpy lobes that have a linear distance-velocity relation will be observed.
This process might account for the formation of bipolar PNe with clumpy lobes, such as NGC 6302.
The energy radiated during the months to years duration of such an event will appear as an intermediate-luminosity optical transient (ILOT)

\end{abstract}

\section{INTRODUCTION}
\label{sec:intro}

Intermediate-Luminosity Optical Transients (\href{http://physics.technion.ac.il/~ILOT}{ILOTs}\footnote{For an updated classification of ILOTs see {http://physics.technion.ac.il/\~ILOT}};
another term in use is Red Novae) are eruptive
stars with peak luminosity between those of novae and supernovae
(e.g. \citealt{Mould1990, Rau2007, Prieto2009, Ofek2008, Botticella2009, Smith2009, Berger2009, Berger2011,
Barbary2009, KulkarniKasliwal2009, Mason2010, Pastorello2010, Kasliwal2011, Tylendaetal2013}).
The typical duration of these eruptions is weeks to years.
The pre-outburst objects of some of the \href{http://physics.technion.ac.il/~ILOT}{ILOTs}, e.g., NGC~300 OT2008-1 (NGC~300OT; \citealt{Bond2009}),
are asymptotic giant branch (AGB) or extreme-AGB stars.
There are single star models (e.g., \citealt{Kochanek2011}) and binary stellar models
\citep{Kashi2010, KashiSoker2010b, SokerKashi2011, SokerKashi2012, SokerKashi2013} for  {ILOT} events harboring AGB stars.

In the binary model a companion star, mostly a main sequence (MS) star, accretes part of the mass ejected by the evolved star.
The gravitational energy released is channelled to radiation, which is behind most of the  {ILOT} brightness,
and mass ejection from the companion, mostly two opposite jets.
More radiation can result from the interaction of the jets with the circumstellar mass (CSM)."
The jets lead to the formation of a bipolar nebula, such as the Homunculus--the bipolar nebula of Eta Carinae \citep{KashiSoker2010a}--that was
formed in the nineteenth century Great Eruption of Eta Carinae.
The connection between the  {ILOT} NGC~300OT and the Great Eruption of $\eta$ Car is discussed in \cite{Kashi2010}.
Other comparisons of  {ILOTs} and LBV eruptions can be found in, e.g., \cite{Smithetal2010, Smithetal2011}, \cite{Humphreysetal2011},
 and \cite{Mauerhan2013} (also \citealt{Levesque2013}).
\cite{SokerKashi2012} further compared the planetary nebulae (PN) NGC~6302 and the pre-PNe OH231.8+4.2, M1-92 and IRAS~22036+5306
with the  {ILOT} NGC~300OT, and proposed that the lobes of some (but not all) PNe and pre-PNe
were formed in a several month-long {ILOT} event (or several close sub-events).
We note that \citet{Prieto2009} already made a connection between the  {ILOT} NGC~300OT and pre-PNe,
and raised the possibility that the progenitor of NGC~300OT was of mass $< 8 ~\rm{M_\odot}$.
Recently more PNe were suggested to have part of their nebula ejected in an  {ILOT} event, e.g., KjPn~8 \citep{BoumisMeaburn2013}.

The nebular part of a PN that is formed by an  {ILOT} event is likely to have the following characteristic properties \citep{SokerKashi2012}.
\begin{enumerate}
\item A linear velocity-distance relation.
This is a consequence of that  {ILOTs} last for time periods $\Delta t_I$ of weeks to several years that is
much shorter than the observed time of hundreds to tens thousands of years later.
In some cases it might be hard to tell whether slower mass elements come from an earlier mass loss episode or
were part of the {ILOT} event but have been slowed down.
\item Bipolar structure.
As the {ILOT} is expected to result from a binary interaction \citep{KashiSoker2010b, SokerKashi2011},
the PN components ejected during the  {ILOT} event(s) are expected to have a bipolar structure.
\cite{SokerKashi2012} argued that most PNe that have been formed by an  {ILOT} event, and hence are bipolar,
are expected to host a binary system with an orbital separation of $\sim 1 \AU$, or
the {ILOT} event took place just as the system entered the common envelope phase. In the later case the orbital separation will
be $\ll 1 \AU$.
\cite{Hajduketal2013} suggest that the 1670 eruption of CK Vul (Nova Vul 1670) that formed a bipolar nebula,
was an  {ILOT} event that was powered by merger event.
\item Expansion velocities of few$\times 100 \km \s^{-1}$.
As we think that most  {ILOTs} are powered by accretion onto a MS star \citep{KashiSoker2010b, SokerKashi2011} that blows jets,
the maximum outflow velocity is similar to that of the escape velocity from MS stars.
The fastest moving elements will be dense parcels of gas that were only slightly slowed-down
while the average velocity of the ejecta will be several times lower because of the interaction with the slower AGB wind.
Therefore, the faster parts of the PN component that was ejected by an  {ILOT} event are expected to
move at velocities of $\sim 100$ -- $1000 \km \s^{-1}$.
\item Total kinetic energy of $\sim 10^{46}$ -- $10^{49} \erg$.
As a typical kinetic energy of  {ILOT} events is in that range,  \cite{SokerKashi2012} argued that the kinetic energy of the ejected
component in a PN is likely to be in that range (see also \citealt{BoumisMeaburn2013}).
\end{enumerate}

The gas in the PN that was not expelled during the  {ILOT} event (or several close events) will not share these properties.
\cite{SokerKashi2012} argued than in some PNe and pre-PNe there are nebular components whose properties are compatible with those listed above, and
hence might hint that these PNe and pre-PNe were shaped by  {ILOT} events.
The nebulae (with the relevant references for their relevant properties) that are listed by \cite{SokerKashi2012} are the PN
NGC~6302 \citep{Meaburn2008, Matsuura2005, Szyszka2011, Wright2011},
and the pre-PNe OH231.8+4.2 \citep{Alcolea2001, Bujarrabal2002, Contreras2004, Kastner1992, Kastner1998},
M1-92 \citep{Bujarrabal1998, Trammell1996, Ueta2007}, and IRAS~22036+5306 \citep{Sahai2003, Sahai2006}.

In this paper we conduct three-dimensional (3D) gas-dynamical simulations to examine the outcome of the interaction of jets
with spherical mass shell ejected by the AGB star very close to the center. The interaction takes place in an optically thick
region, such that the photon-diffusion time is longer than the relevant flow time (section \ref{sec:diffusion}).
In section \ref{sec:numerical} we describe the numerical set-up, and in section \ref{sec:results} we present our numerical results.
Our summary is in section \ref{sec:summary}.

\section{THE OPTICALLY THICK REGIME}
\label{sec:diffusion}

When interaction occurs at a large distance from the center the density of the post-shock jets' material is sufficiently low
that radiative cooling occurs over a longer than the flow time, $t_{\rm rad} \ga t_f$.
This type of interaction is referred to also as energy conserving.
Closer to the center, and when the region is optically thin, the radiative cooling time is short and
a substantial fraction of the thermal energy of the post-shock jets' material is radiated away.
This is referred to also as momentum conserving interaction.
These two regimes are assumed to occur in most PNe, and most numerical simulations of PN jet-shaping deal with one of these
two regimes (e.g., \citealt{LeeSahai2003, LeeSahai2004, Akashietal2008, AkashiSoker2008, Dennisetal2008, Dennisetal2009, Leeetal2009, Huarte-Espinosa2012};  Balick et al. 2013 ).
We note that \cite{GarciaFrank2004} simulate interaction of jets with the AGB wind very close to the binary system, but the mass loss rates,
hence densities, were low and the interaction region was optically thin.
In cases where the interaction occurs close to the center and mass loss rates are high, the interaction region can be optically thick.
This type of flow will be studied here.

The flow time can be defined as the acceleration time of the outer shell, or approximately $r/v_{\rm f}$,
where the flow speed $v_f$ is taken to be the final shell speed.
In cases studied here the jets are relatively massive, and the average velocity of the shell can be
$v_f \simeq$few$\times 100 \km \s^{-1}$. We therefore scale the flow time with
\begin{eqnarray}
t_{\rm f} \equiv \frac{r}{v_f} \simeq 50
\left( \frac {r}{1000 \AU} \right)
\left( \frac {v_f}{100 \km \s^{-1}} \right)^{-1} \yr
\label{eq:tflow}
\end{eqnarray}
The jets' post shock temperature is
\begin{eqnarray}
T_p \simeq 1.4 \times 10^7
\left( \frac {v_j}{1000 \km \s^{-1}} \right)^{2} \K .
\label{eq:temp}
\end{eqnarray}
The cooling function at that temperature and below is \citep{Gaetz88}
$\Lambda \simeq 10^{-22} (T/10^6 \K)^{-1/2} \erg \s^{-1} \cm^3$, such that the cooling time at a constant pressure
is
\begin{eqnarray}
t_{\rm rad} = \frac{5}{2} \frac {nkT}{n_e n_p \Lambda} \simeq 60
\left( \frac {v_j}{1000 \km \s^{-1}} \right)^{4}
\left( \frac {\dot M_f}{10^{-4} M_\odot \yr^{-1}} \right)^{-1}
\left( \frac {\delta}{0.2} \right)
\left( \frac {r}{1000 \AU} \right)^2 \yr,
\label{eq:tcool1}
\end{eqnarray}
where $\rho=4 \rho_j$ has been used for the jets' post-shock density, with the density of the pre-shock
gas being
\begin{equation}
\rho_j =  \frac {\dot M_{\rm 2jets}}{4 \pi \delta r^2 v_j}.
\label{eq:rhof}
\end{equation}
Here $\dot M_{\rm 2jets}$ is the jets' mass outflow rate in both directions, and $\delta$ is defined such that
the two opposite jets cover in total a solid angle of $4 \pi \delta$.
The adiabatic case, where $t_{\rm rad} > t_{\rm f}$, occurs for interaction distance from the center of
\begin{eqnarray}
r_{\rm ad} \ga 1000
\left( \frac {\dot M_f}{10^{-4} M_\odot \yr^{-1}} \right)
\left( \frac {v_j}{1000 \km \s^{-1}} \right)^{-4}
\left( \frac {v_f}{100 \km \s^{-1}} \right)^{-1}
\left( \frac {\delta}{0.2} \right)^{-1} \AU
\label{eq:radiabatic}
\end{eqnarray}

Equation \ref{eq:tcool1} and \ref{eq:radiabatic} assume that the radiative cooling occur in an optically thin region.
This is not the case very close to the center and for very high mass loss rates, as the cases we deal with here.
For electron scattering opacity of $\kappa=0.34$ the approximate optical depth in the radial direction is
$\tau \simeq \kappa \rho r$.
For an  {ILOT}
 event of mass loss from an AGB star we take the radius to be the  {ILOTs} duration times the
AGB wind speed $r_I \simeq \Delta t_I v_{\rm AGB} \sim (1-5) \yr \times (10-30) \km \s^{-1} \simeq 10 \AU$.
The total mass $M_I$ inside this radius is the mass ejected by the AGB and that ejected by the companion in jets.
The optical depth along the radial direction is
\begin{eqnarray}
\tau_r  \simeq \frac {M_I}{4 \pi r_I^2} \kappa \simeq 240
\left( \frac {M_I}{0.1M_\odot} \right)
\left( \frac {r_I}{10 \AU} \right)^{-2} .
\label{eq:taur}
\end{eqnarray}
The diffusion time \citep{Arnett1979} is
\begin{eqnarray}
\tau_{\rm diff} = \frac{M \kappa}{ 4 r_I c }
\simeq 0.12
\left( \frac {M_I}{0.1M_\odot} \right)
\left( \frac {r_I}{10 \AU} \right)^{-1} \yr,
\label{eq:tdiff1}
\end{eqnarray}
where c is the light speed.
The energy decreases according to $d \ln E /dt = - \tau_{\rm diff}^{-1}$ (neglecting heating and adiabatic cooling).
In this type of interaction the mass in the jets is not negligible relative to the mass ejected by the AGB, and the final outflow velocity will be
several$\times 100 \km \s^{-1}$.
The flow time is $t_f=r_I/v_f$, and hence the ratio of diffusion time to flow time $r_I/v_f$ is
\begin{eqnarray}
\frac {\tau_{\rm diff}} {t_{\rm f}} =
\frac{M \kappa}{ 4 r_I^2  } \frac{v_f}{c} = \pi \tau_r \frac{v_f}{c}
\simeq 1.3
\left( \frac {M_I}{0.1M_\odot} \right)
\left( \frac {v_f}{500 \km \s^{-1}} \right)
\left( \frac {r_I}{10 \AU} \right)^{-2} .
\label{eq:tfdiff}
\end{eqnarray}
This ratio implies that the gas will not cool as rapidly as it radiates locally, as photons diffusion time is not much shorter than the flow time.
As well, instabilities, such as Rayleigh-Taylor instabilities, on scales of $\sim 0.1 r_I$ that are known to exist in energy conserving flows,
will develop on time scales much shorter than the flow time, and hence much shorter than the diffusion time scale.
We now turn to simulate such cases.

\section{NUMERICAL SETUP}
\label{sec:numerical}

Our simulations are performed by using version 4.0-beta of the FLASH code \citep{Fryxell2000}.
The FLASH code is an  adaptive-mesh refinement modular code used for solving
hydrodynamics or magnetohydrodynamics problems.
Here we use the unsplit PPM (piecewise-parabolic method) solver of FLASH.
We neither include gravity nor radiative cooling as the interaction region is optically thick (section \ref{sec:diffusion}).
Instead of calculating radiative cooling and radiative transfer, that are too complicated for the flow geometry,
we lower the adiabatic index $\gamma$ to mimic cooling by photon diffusion.
If half the energy is lost, it is as if there are three internal degree of freedom (total of 6 degrees of freedom),
and we take $\gamma=8/6=1.33$.
For $\tau_{\rm diff}/{t_{\rm f}} =0.25$, for example $\sim 80\%$ of the energy is lost in radiation.
This can be simulated with $\gamma=1.13$ (15 degrees of freedom).
Based on this consideration we will simulate several values of the adiabatic index $\gamma$.
Using lower values of $\gamma$ to mimic radiative cooling is reasonable when kinetic energy is channelled to
thermal energy, but not when thermal energy is channelled to kinetic energy.
In this study we focus only on the early stages of the interaction, so this approximation is adequate.

We employ a full 3D adaptive mesh refinement (AMR, 7 levels; $2^{10}$ cells in each direction) using
a Cartesian grid $(x,y,z)$ with outflow boundary conditions at all boundary surfaces.
We define the $x-y$ ($z=0$) plane to be the equatorial plane of the PN and simulate the whole space (the two sides of
the equatorial plane).
The grid size is $10^{15} \cm $ in the $x$ and $y$ directions, and $ 2 \times 10^{15} \cm $ in the $z$ direction.
{{{The number of cells in all directions (x,y,z) is the same, and each grid cell has its $z$ axis twice as long as its $x$ or $y$ axis.  }}}

At $t=0$ we place a spherical dense shell in the region $R_{\rm in} = 10^{14} \cm < r < 2\times 10^{14} \cm= R_{\rm out}$,
and with a density profile of
$\rho_s = 1.58 \times 10^{-11}(r/10^{14} \cm)^{-2} \g \cm^{-3}$, such that the total mass in the shell is $0.1M_\odot$.
The gas in the shell has an initial radial velocity of $v_s = 10 \km \s^{-1}$.
The shell corresponds to a mass loss episode lasting for $\sim 3 \yr$ and with a constant mass loss rate of $\dot M_s \simeq 0.03 M_\odot \yr^{-1}$.
The regions outside and inside the dense shell are filled with a lower density spherically-symmetric slow wind
having a uniform radial velocity of  $v_{\rm wind}=v_s= 10 \km \s^{-1}$.
The slow wind density at $t=0$ is taken to be $\rho(t=0) = {\dot M_{\rm wind}}({4 \pi r^{2} v})^{-1}$,
where $\dot M_{\rm wind}=10^{-5} {\rm M_\odot \yr^{-1}}$.

\begin{figure}
\centering
\includegraphics[scale=0.7]{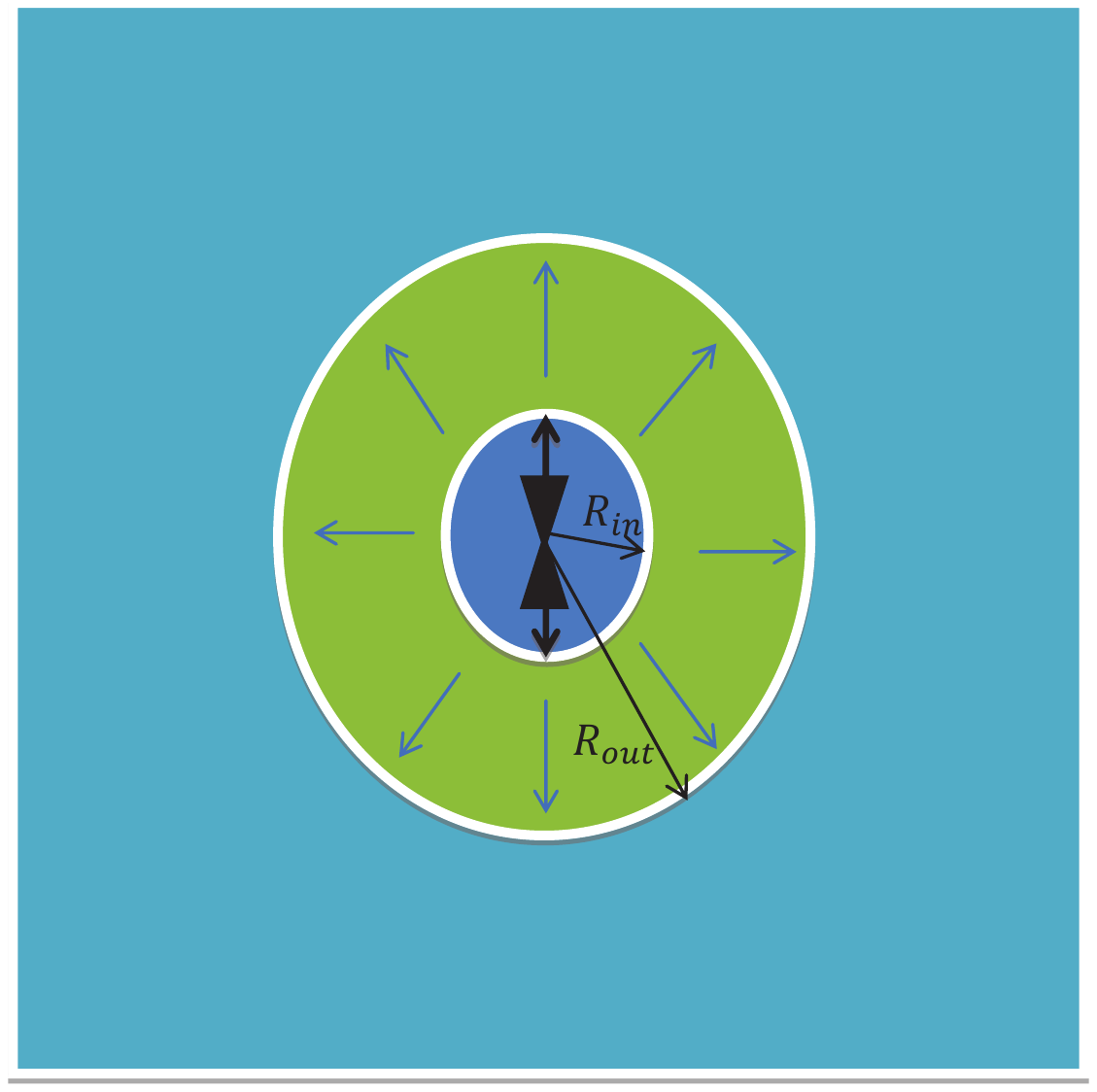}
\caption{The flow set-up. The two opposite jets are launched from near the center starting at $t=0$, and are active for two months.
  A spherical dense shell with an outward velocity of $v_s = 10 \km \s^{-1}$ is placed in the region $10^{14} \cm < r < 2\times 10^{14} \cm$.
  The regions not occupied by the dense shell are filled at $t=0$ with a low density wind radially expanding with a velocity of $v_{\rm wind} =v_s$. }
\label{fig:scheme0}
\end{figure}

The two opposite jets are lunched from the inner $5 \times 10^{13} \cm $ region and within a half opening angle of
$\alpha = 50^\circ$ ($0<\theta<\alpha$) along the $z-$axis.
By the term `jets' we refer also to wide outflows, as we simulate here. More generally, we simulate slow-massive-wide (SMW) outflows.
The launching episode lasts for $5 \times 10^{6} \s = 58 \days$.
The jets' initial velocity is $v_{\rm jet}=1000 \km \s^{-1}$, and the mass loss rate into the two jets together
is $\dot M_{\rm 2jets} = 0.13 M_\odot \yr^{-1}$. 
The slow wind, dense shell, and the ejected jets start with a temperature of $1000 \K$.
The initial jets' temperature has no influence on the results (as long it is highly supersonic) because the jets
rapidly cool due to adiabatic expansion.
The adiabatic index $\gamma$ will be varied between different runs.
For numerical reasons a weak slow wind is injected in the sector $\alpha<\theta<90^\circ$.
{{{No seed perturbations are introduced in the initial conditions of our simulations. The limited numerical resolution supplies the seeds of the instabilities.}}}
The initial flow structure is depicted in Figure \ref{fig:scheme0}.

Our calculations do not include the ionizing radiation and the fast wind blown by the central star during the PN phase.
We simply aim at showing the shaping of the nebular gas during the vigourous jet-shell interaction of an  {ILOT} event.
As well, we don't try to reproduce the structure of NGC~6302 as it evolves several nebular segments that were not formed
in an  {ILOT} event. We rather limit ourself to produce a very clumpy lobes with a distance-velocity linear relation.
The mass of $\sim 0.1-1M_\odot$ and velocity range of $0-500 \km \s^{-1}$ in the lobes of NGC~6302
serve as guiding properties.

\section{RESULTS}
\label{sec:results}

\begin{figure}
\centering
\includegraphics[trim=4cm 12cm 3cm 1.2cm, clip=true, scale=0.7]{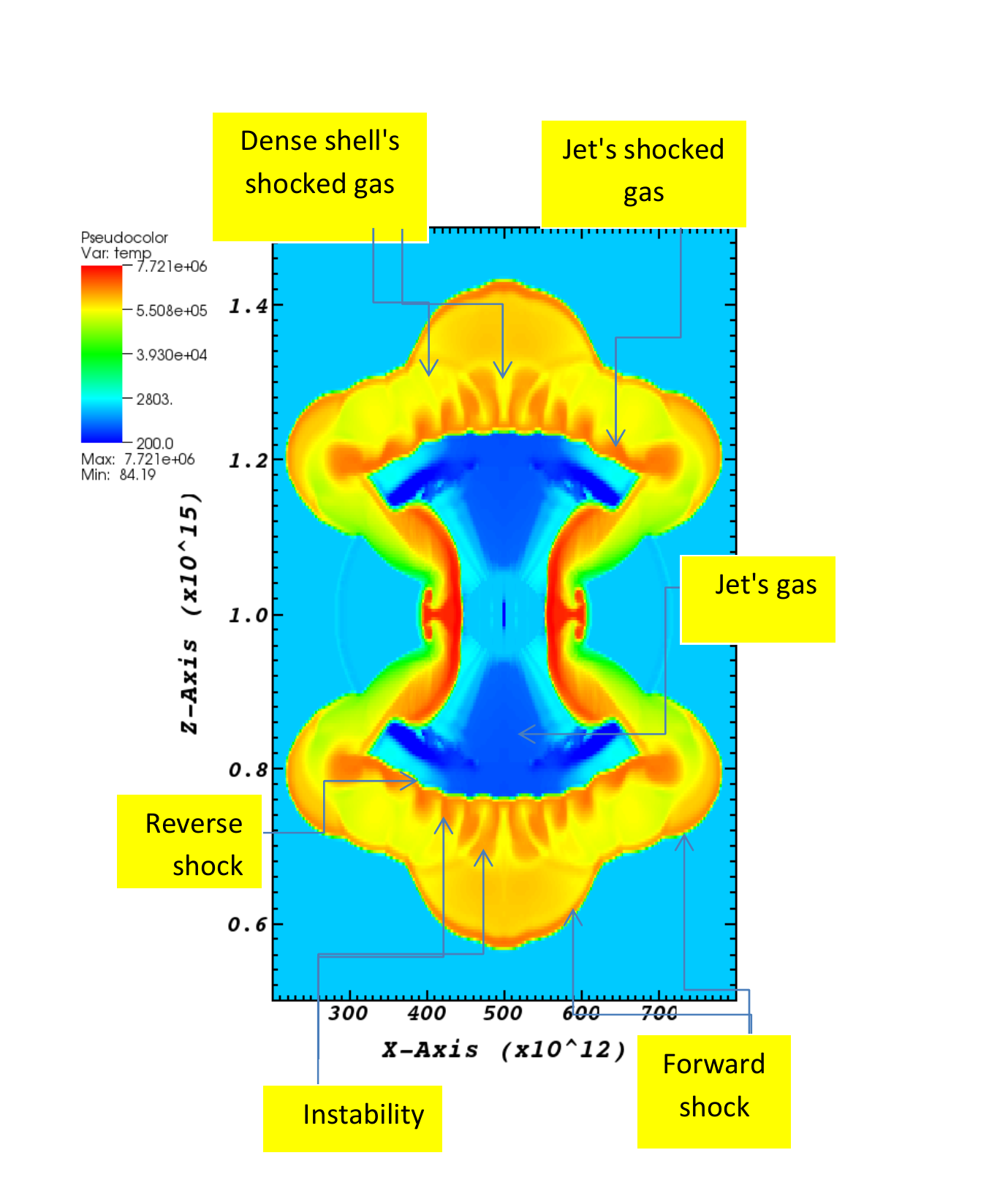}
    \caption{The temperature map for the run with $\gamma=1.1$ at $t=76.1 \days$, in logarithmic scale. The color bar is in $K$ and the units along the axes are in $\cm$.
     Several flow components are marked on the figure.
    We simulate the entire space and apply no symmetry-folding.
    The density map is given in Fig. \ref{fig:dens11}.  }
\label{fig:General}
\end{figure}

\begin{figure}
\hfill
\subfigure[$t=34.5$ days]{\includegraphics[trim=1.2cm 1cm 1.9cm 0.1cm,clip=true, width=5.cm]{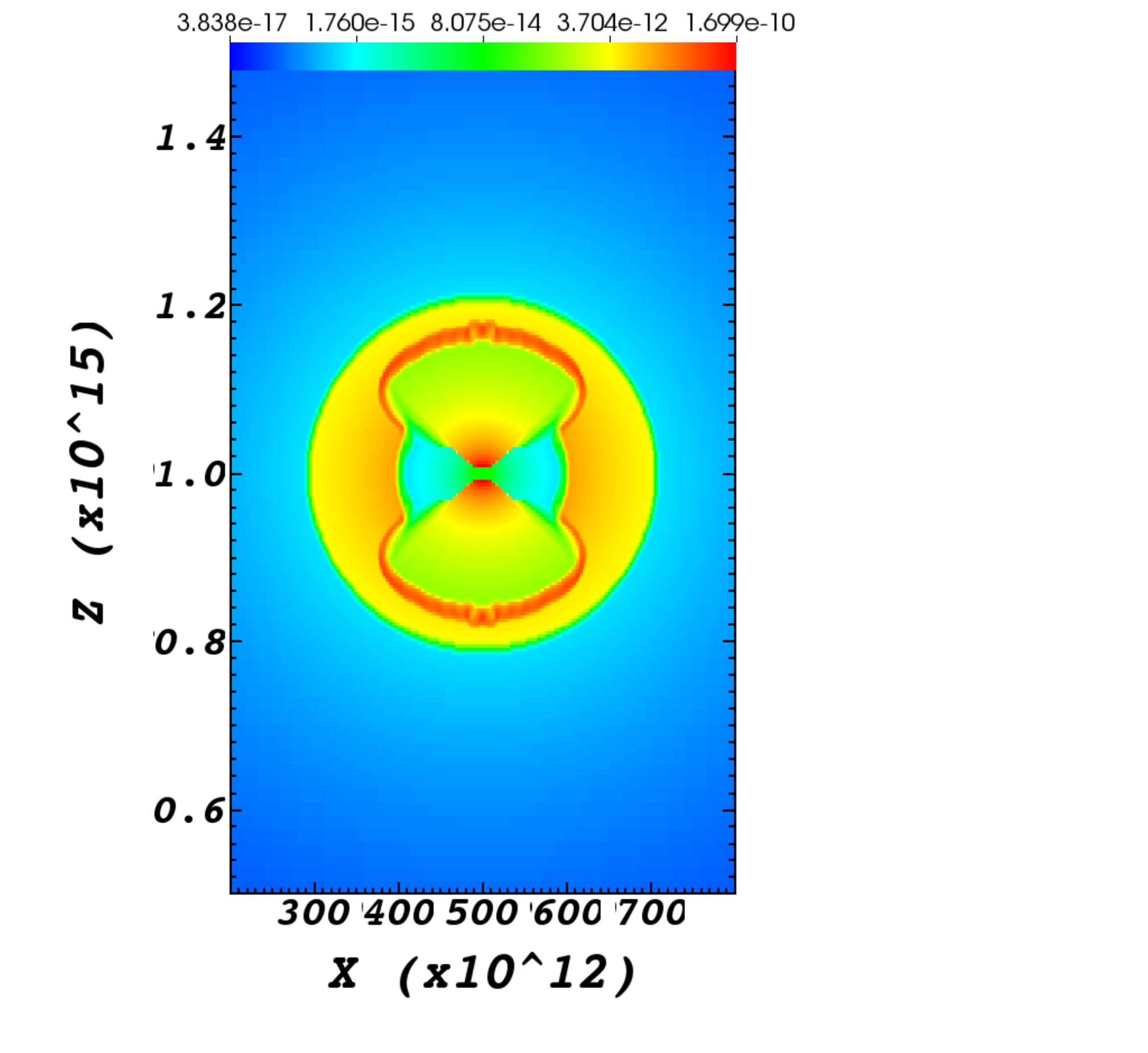}}
\hfill
\subfigure[$t=57.6$ days]{\includegraphics[trim=1.2cm 1cm 1.9cm 0.1cm,clip=true, width=5.cm]{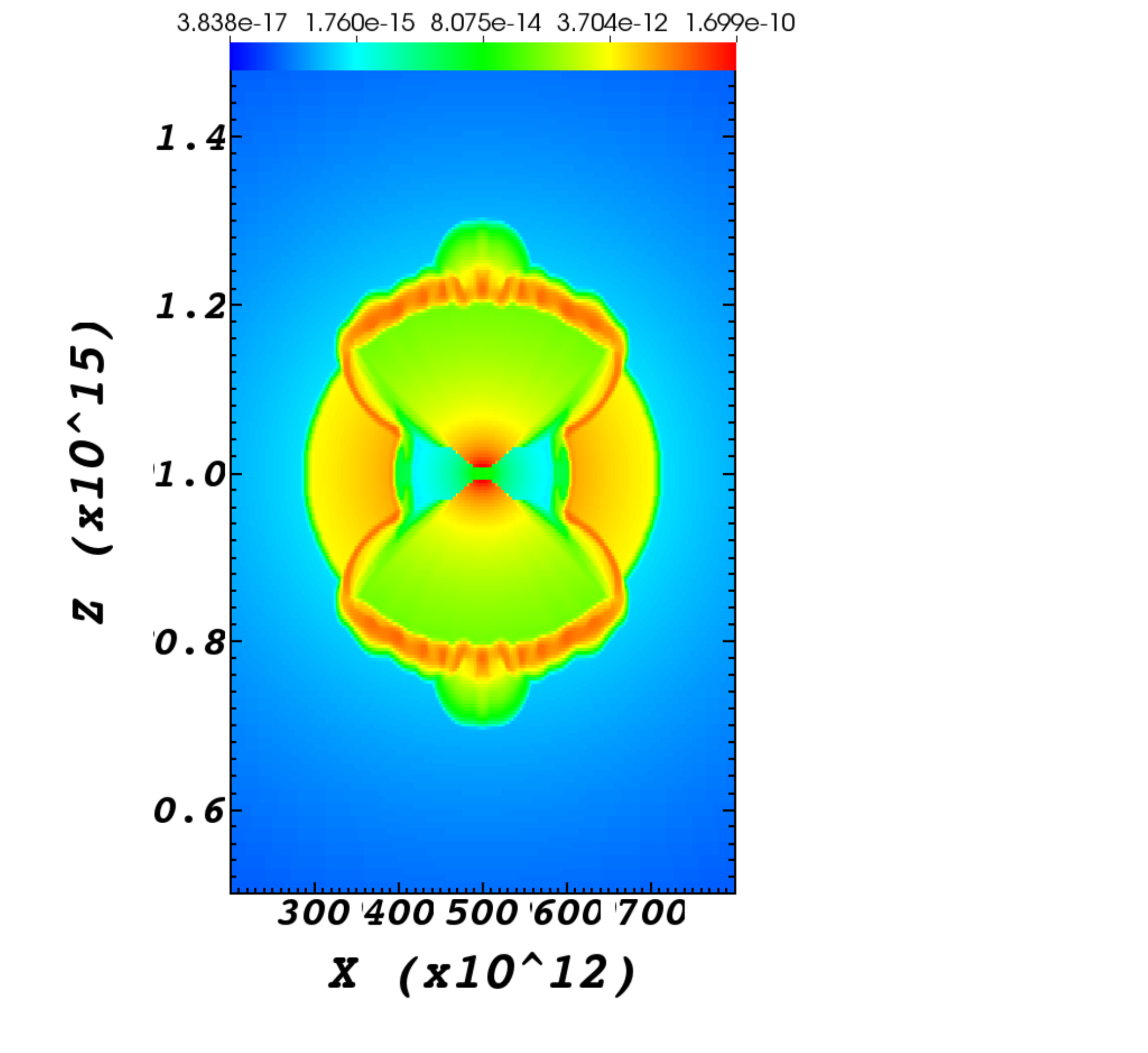}}
\hfill
\subfigure[$t=76.1$ days]{\includegraphics[trim=1.2cm 1cm 1.9cm 0.1cm,clip=true, width=5.cm]{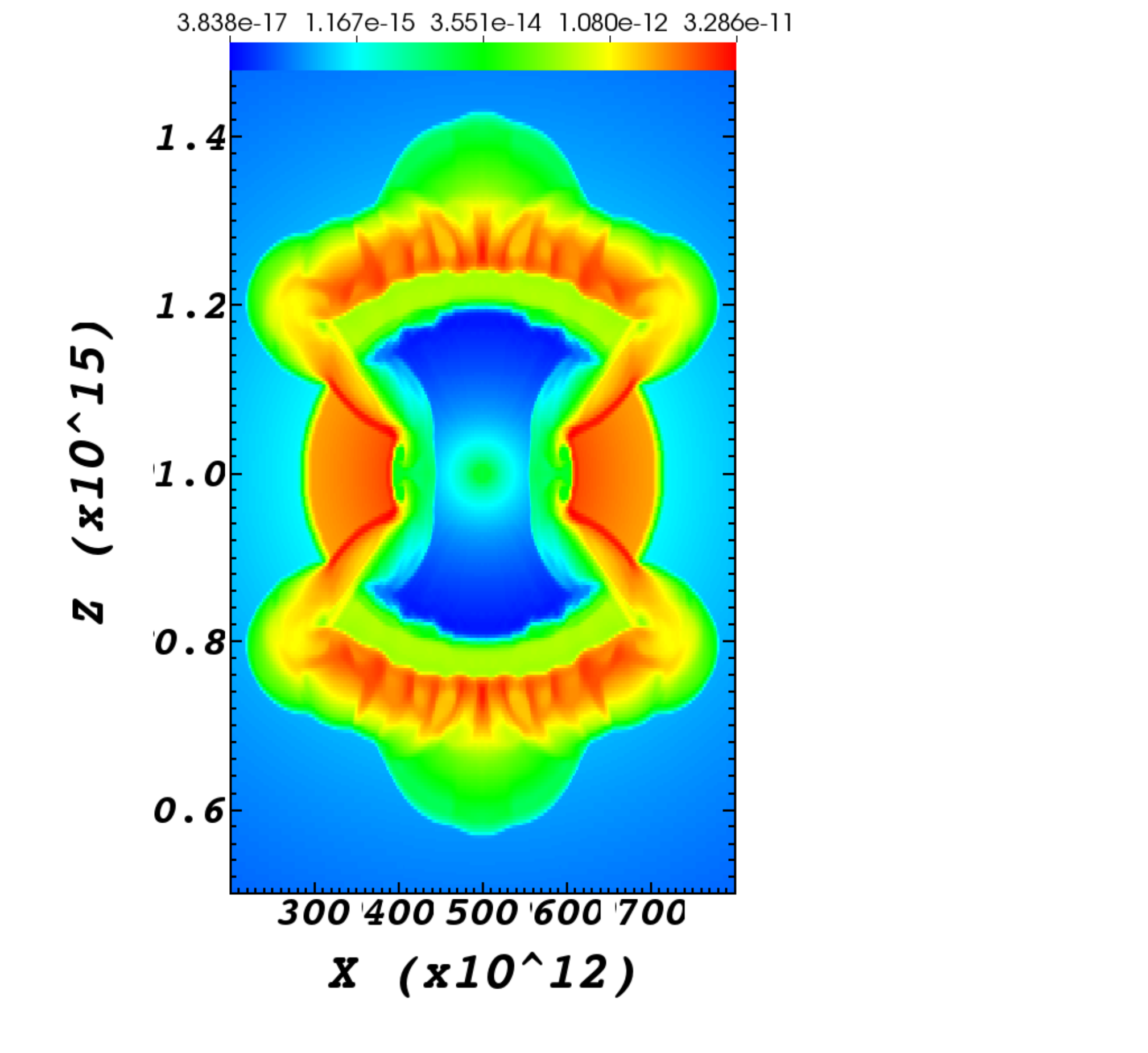}}
\hfill
\caption{The density maps at three times of the $\gamma=1.1$ run, in logarithmic scale.
Color coding is in $\g \cm^{-3}$.
The times of panels a, b, and c, are $34.5$,  $57.6$, and $76.1$ days, respectively.
The third panel corresponds to the temperature map in Fig. \ref{fig:General}.
{{{Note that the color logarithmic scales in the bars are not identical in the different panels}}}.
Units on the axes are in $\cm$.}
\label{fig:dens11}
\end{figure}

We here present the results of the 3D simulations of the flow structure during an  {ILOT} event.
We assume that the AGB star ejected a shell within few years and the companion launched two jets for a period of two months.
Following the flow for a longer time is numerically challenging, and is postponed for a future study.
We first describe the results of a run with adiabatic index of $\gamma=1.1$.
In Fig. \ref{fig:General} we introduce some features of the flow by presenting the temperature map at $t=76.1 \days$.
The temperature scale is given in the bar on the left in units of $K$.
The density maps at three times of the same run are shown in Fig.\ref{fig:dens11}.

The prominent features seen in Fig. \ref{fig:General} are as follows.
The freely expanding jets' material encounters a shock, termed reverse shock, and is heated to $\sim 10^7 \K$.
For lower values of $\gamma$ the postshock temperature will be lower.
Due to expansion the shocked jets' material suffers adiabatic cooling.
A forward shock is running into the dense shell, and later into the slow wind.
At the time presented in Fig. \ref{fig:General} the forward shock is already running into the slow wind.
The two lobes form a bipolar nebula, with two opposite protrusions along the $z-$axis, quite similar to those
seen in the PN Mz-3 (e.g., \citealt{Guerreroetal2004}).
Also seen is a concentration of dense gas in the equatorial plane and close to the central star ($r=7 \times 10^{13} \cm$).
This gas might be observed later as a ring in the equatorial plane.
Instability fingers (tongues) are seen in the interface between the shocked jets' and shell media.
They appear early on, but become prominent only after $\sim 2$~months.

The post-shock jets' material has a density of $\sim 1 \times 10^{-11} \g \cm^{-3}$,
which is about five times as low as the density of the post-shock dense-shell gas.
The post-shock jets' material accelerates the denser shocked shell gas, and the flow becomes Rayleigh-Taylor (RT) unstable.
The resulting instability fingers are clearly seen in Fig. \ref{fig:General} and in later figures.
Thin-shell instabilities might also be involved as the structure of the developed instabilities resemble
that of the non-linear thin-shell instability simulated by \cite{McLeod2013}.

\begin{figure}
\centering
\subfigure[$t=34.5$ days]{\includegraphics[trim=4cm 1cm 2.5cm 1.2cm, clip=true, width=5.cm]{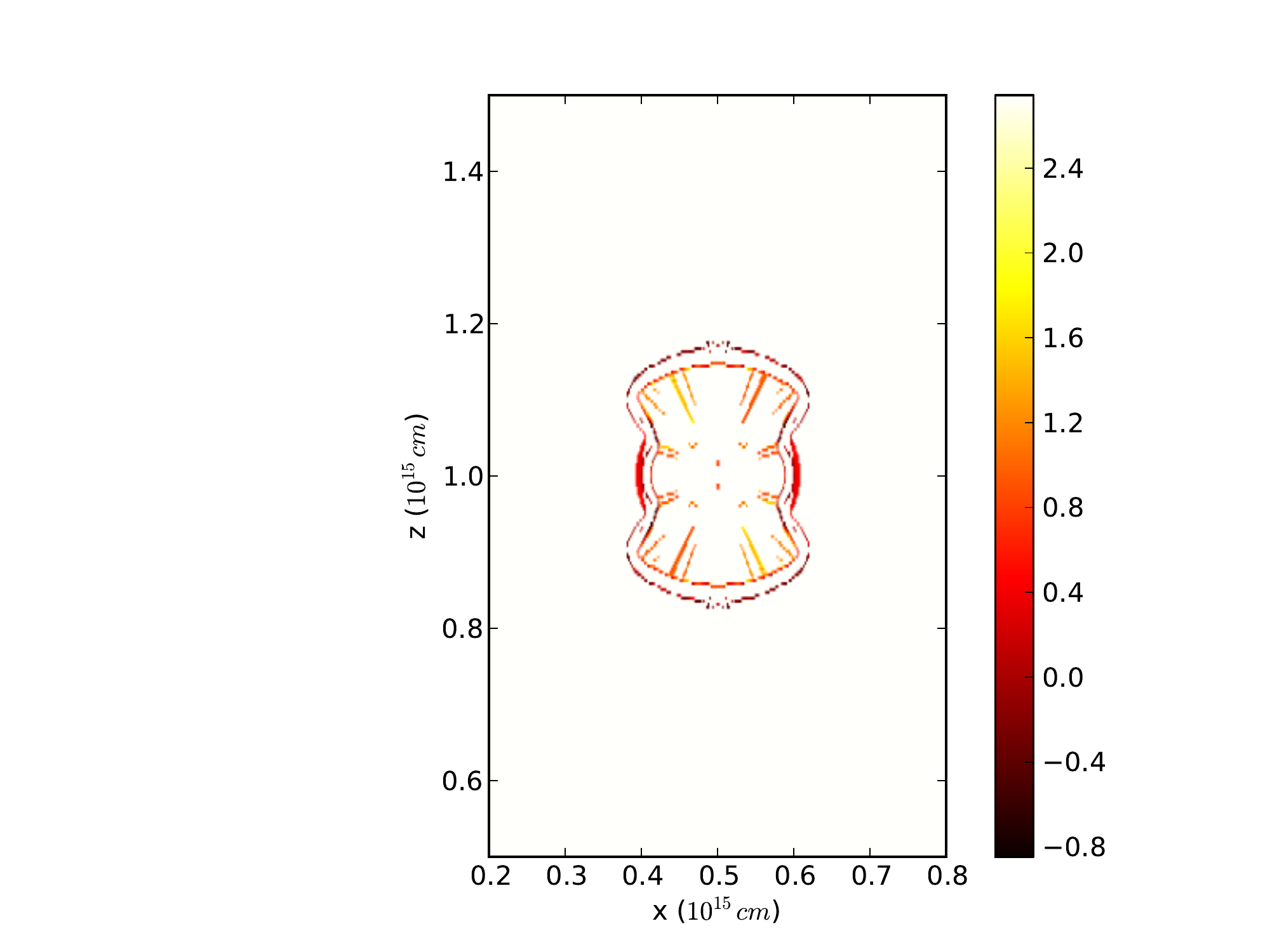}}
\centering
\subfigure[$t=57.6$ days]{\includegraphics[trim=4cm 1cm 2.cm 1.2cm, clip=true, width=4.75cm]{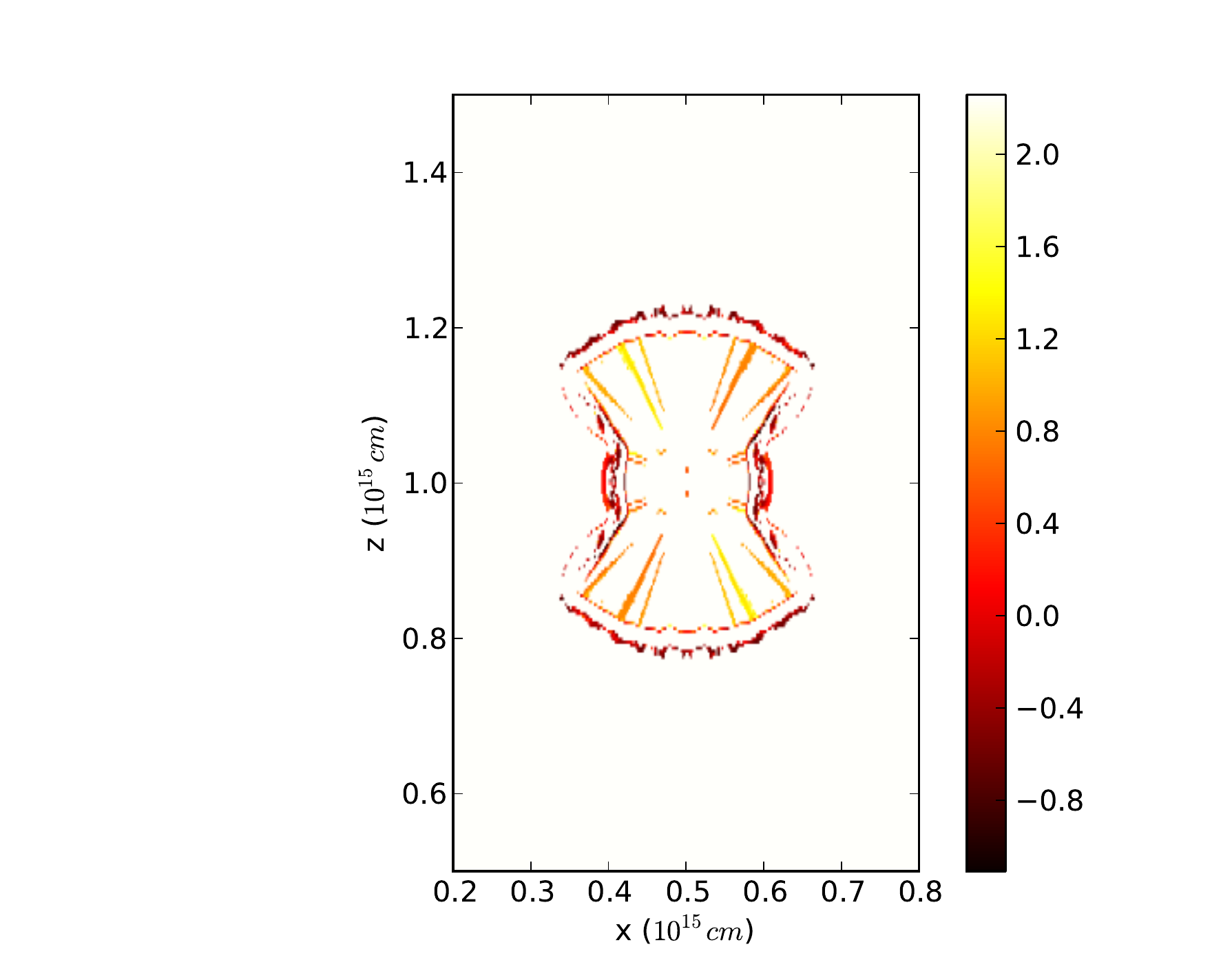}}
\centering
\subfigure[$t=76.1$ days]{\includegraphics[trim=4cm 1cm 2.5cm 1.2cm, clip=true, width=5.cm]{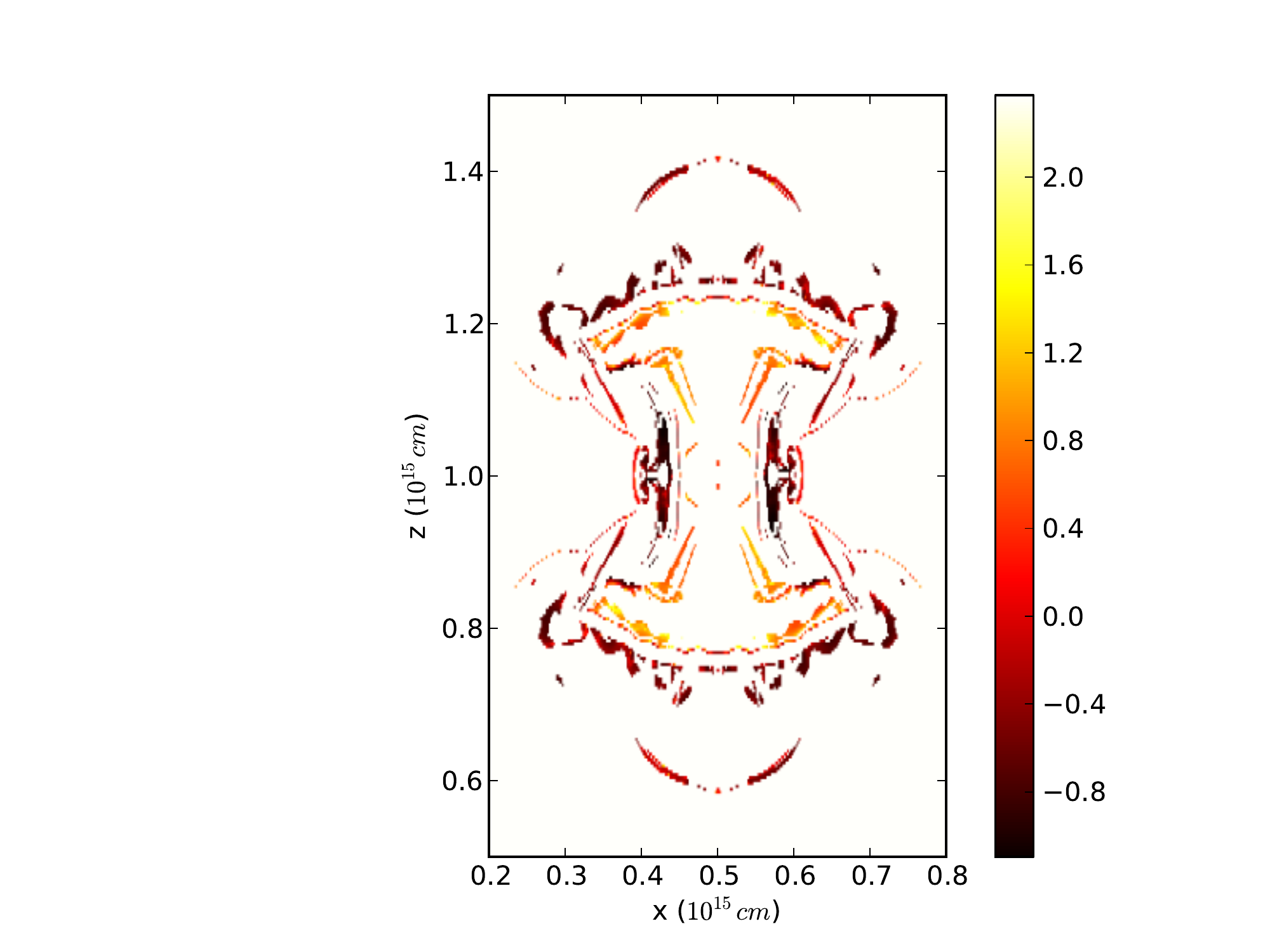}}
\hfill
\caption{The ratio of the Rayleigh-Taylor instability growth time to the time of the simulation according to equation (\ref{eq:rt}), for the run with $\gamma=1.1$, in logarithmic scale.
Red regions are less stable. In yellow regions the instability growth time is long. White regions are stable. {{{The white regions have growth time that is practically infinite, hence coincide with stable regions. Note that the color scale is not identical in the different panels.}}}
The times of panels a, b, and c, are $34.5$,  $57.6$, and $76.1$ days, respectively.
Units on the axes are in $ 10^{15} \cm$.}
  \label{fig:RT}
\end{figure}

In Fig. \ref{fig:RT} we show the ratio of the RT-growth time $\tau_{\rm RT}$, to the present time of the simulation $t$
for the run with $\gamma=1.1$.
This ratio is shown only in regions that are RT-unstable by the condition $\nabla{P} \cdot \nabla{\rho}<0$, and is
calculated as
\begin{equation}
\frac{\tau_{\rm RT}}{t} = \frac{1}{t} \sqrt{\frac{\lambda \rho}{\vert \vec{\nabla} P \vert}},
\label{eq:rt}
\end{equation}
where $\lambda$ is the typical size of the RT instabilities and $\vec{\nabla} P$ is the pressure gradient.
We take $\lambda$ to be, somewhat arbitrarily, $10^{13} \cm$,
The exact value is of no significance for our analysis.
Figure \ref{fig:RT} emphasizes the regions that are RT-unstable, from which the development of the instability fingers seen in figures
\ref{fig:General} and \ref{fig:dens11} can be understood.

\begin{figure}
\centering
\subfigure[Density with arrows]{\includegraphics[width=11cm]{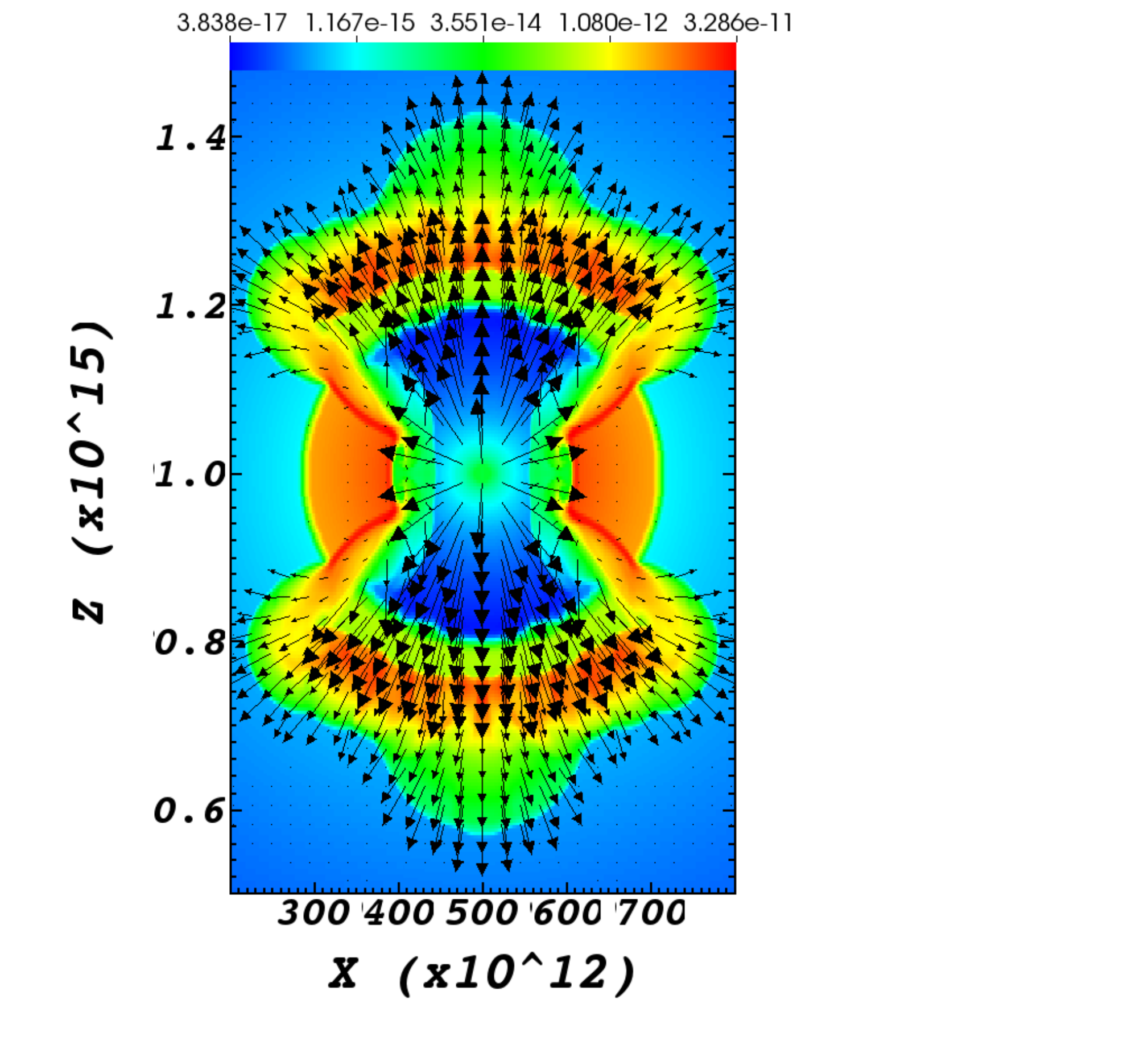}}
\centering
\subfigure[Velocity magnitude]{\includegraphics[width=11cm]{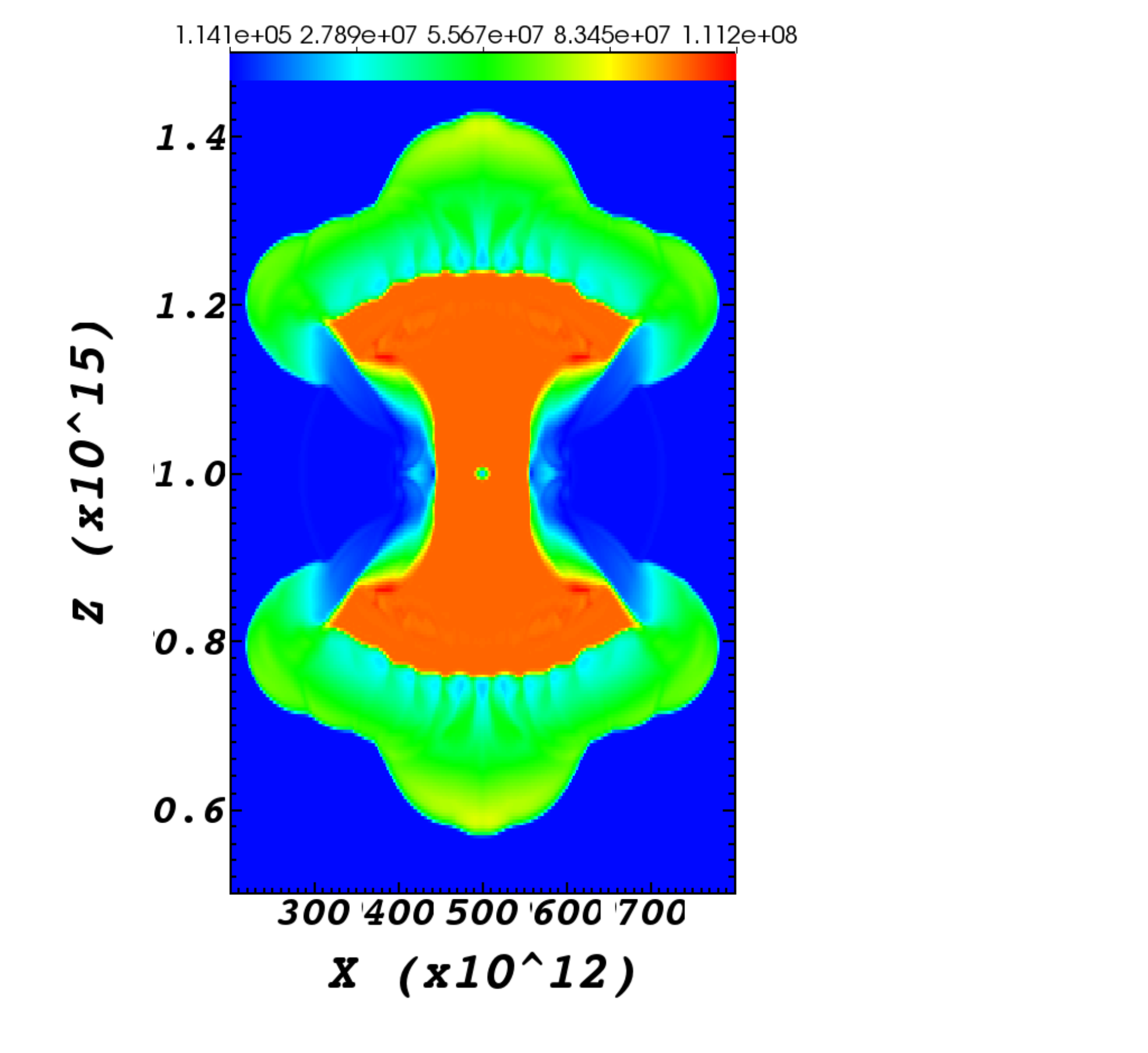}}
\caption{Maps of the density and velocity structure and of the velocity magnitude for the run with $\gamma=1.1$ at
$t =76.1 \days$ (corresponding to figures \ref {fig:General}, \ref{fig:dens11}c, and \ref{fig:RT}c), in logarithmic scale.
Units of density, velocity, and axes are $ \g \cm^{-3}$, $\cm \s^{-1}$, and $\cm$, respectively.}
  \label{fig:dvt}
\end{figure}

\begin{figure}
\centering
\includegraphics[scale=0.55]{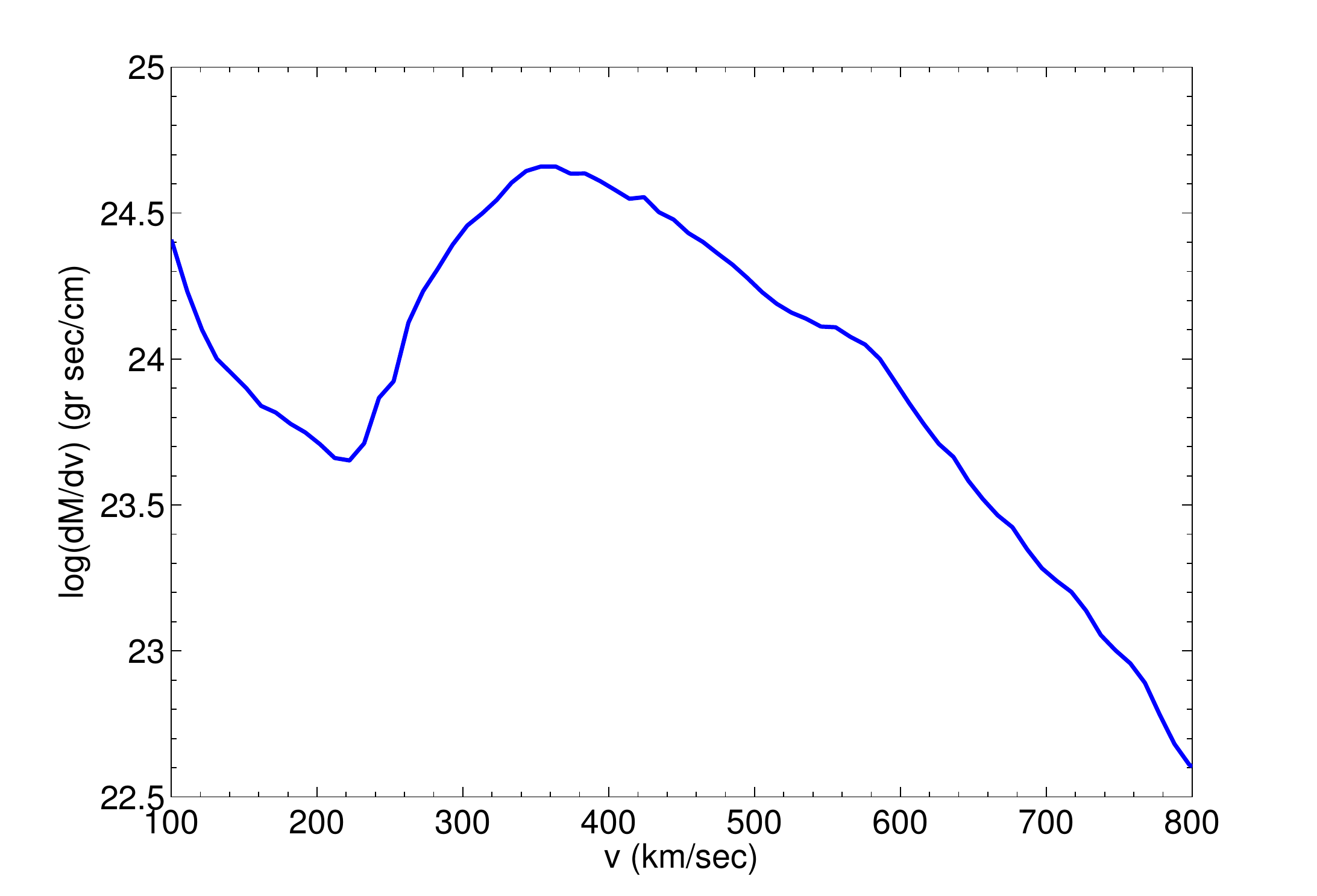}
   \caption{The mass distribution as a function of the velocity magnitude for the $\gamma=1.1$ run. }
\label{fig:dMdv}
\end{figure}

To close our study of the flow properties for the $\gamma=1.1$ run, in Fig. \ref{fig:dvt} we present the velocity map and magnitude,
and in Fig. \ref{fig:dMdv} we show the nebular mass distribution ${dM}/{dv}$ as function of velocity $v$
(the magnitude of the velocity vector in each cell). From these figures we learn that most of the slow gas, $v \la 50 \km \s^{-1}$
resides in the equatorial plane and the regions further out.
In the lobes most of the mass has velocities in the range $250 \km \s^{-1} \la  v_{\rm lobe} \la 600 \km \s^{-1}$.

\begin{figure}
\centering
\subfigure[$\gamma=1.02$]{\includegraphics[trim=1.2cm 1cm 1.9cm 0.1cm,clip=true, width=5.cm]{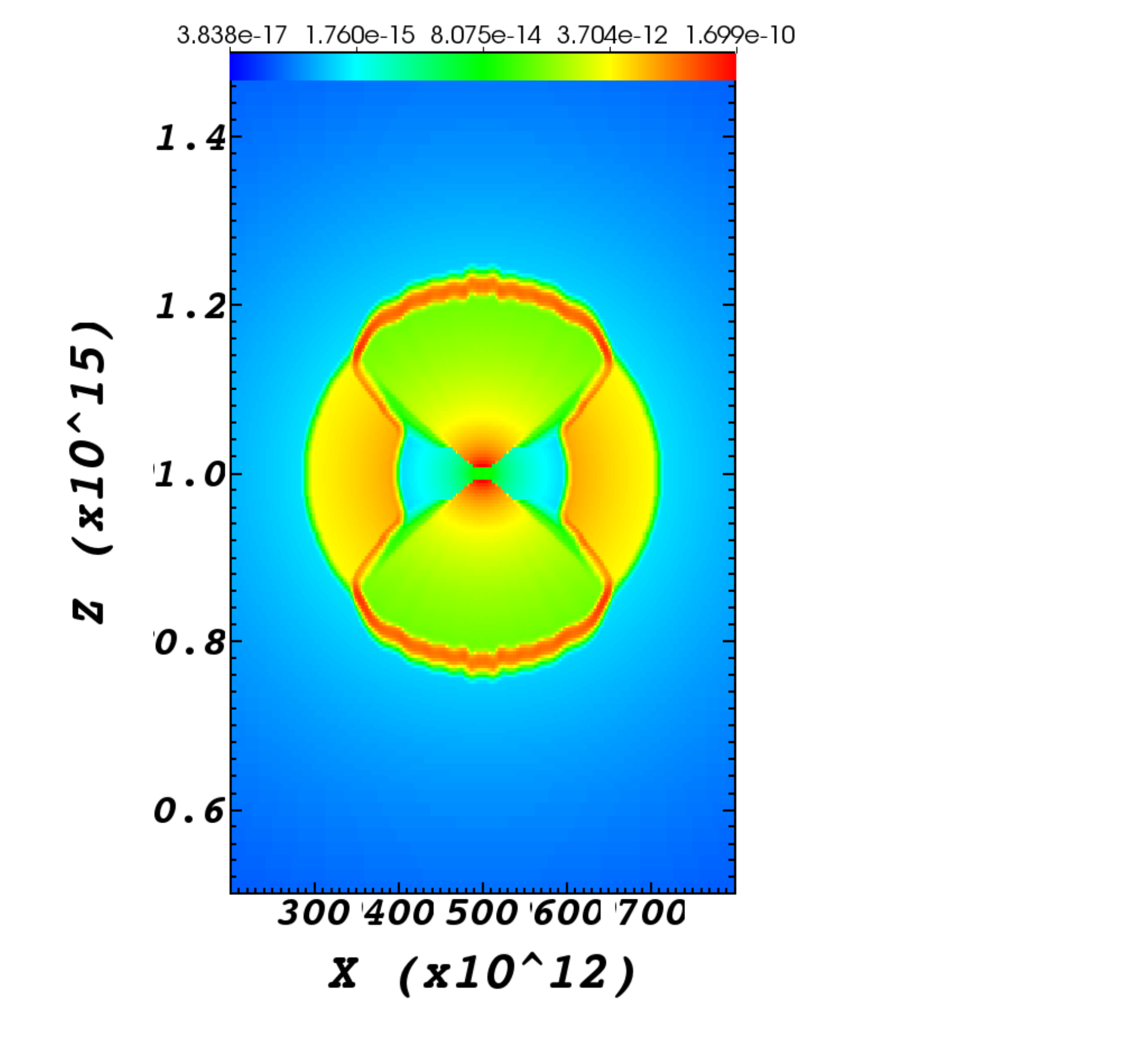}}
\centering
\subfigure[$\gamma=1.05$]{\includegraphics[trim=1.2cm 1cm 1.9cm 0.1cm,clip=true, width=5.cm]{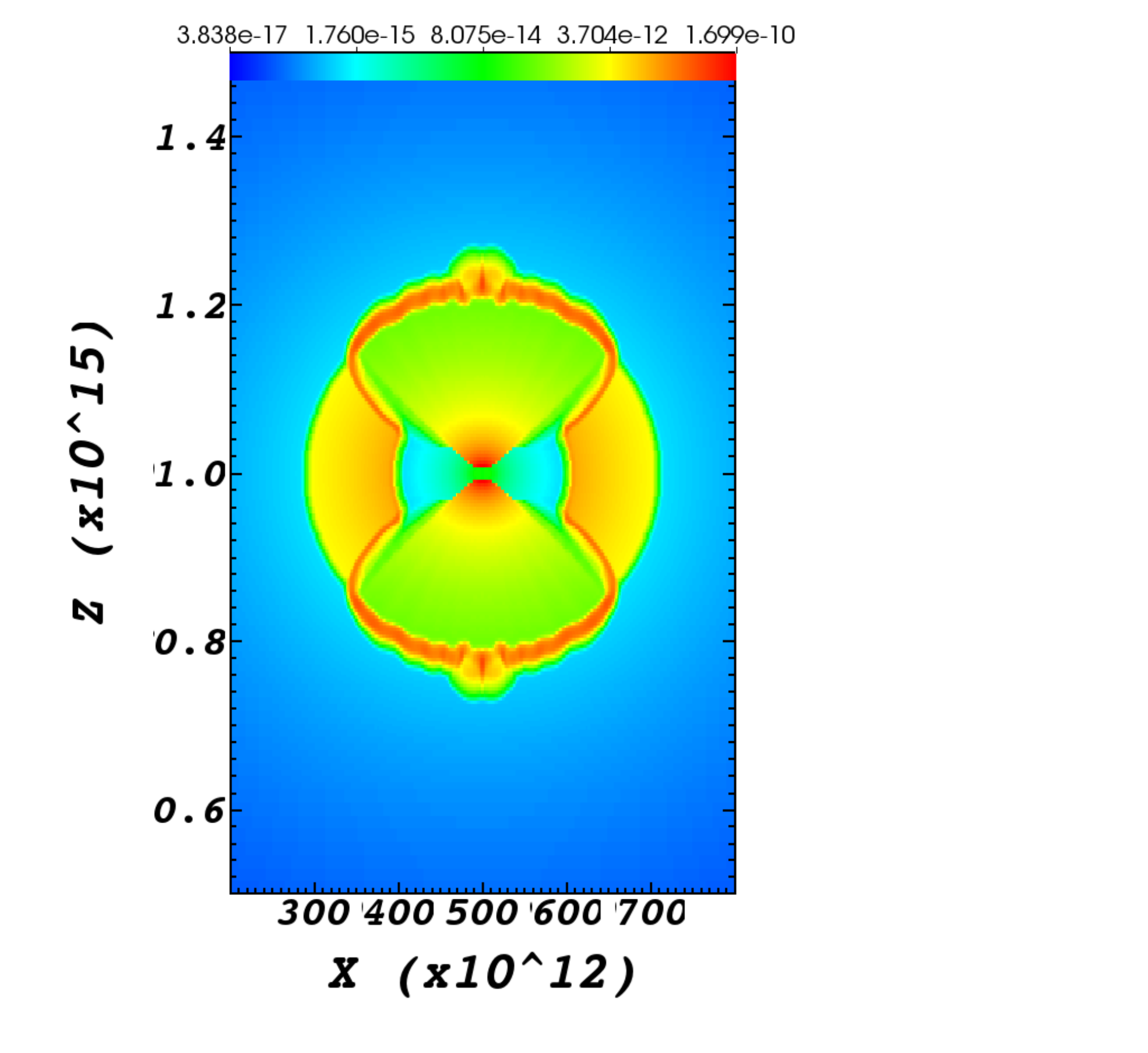}}
\centering
\subfigure[$\gamma=5/3$]{\includegraphics[trim=1.2cm 1cm 1.9cm 0.1cm,clip=true, width=5.cm]{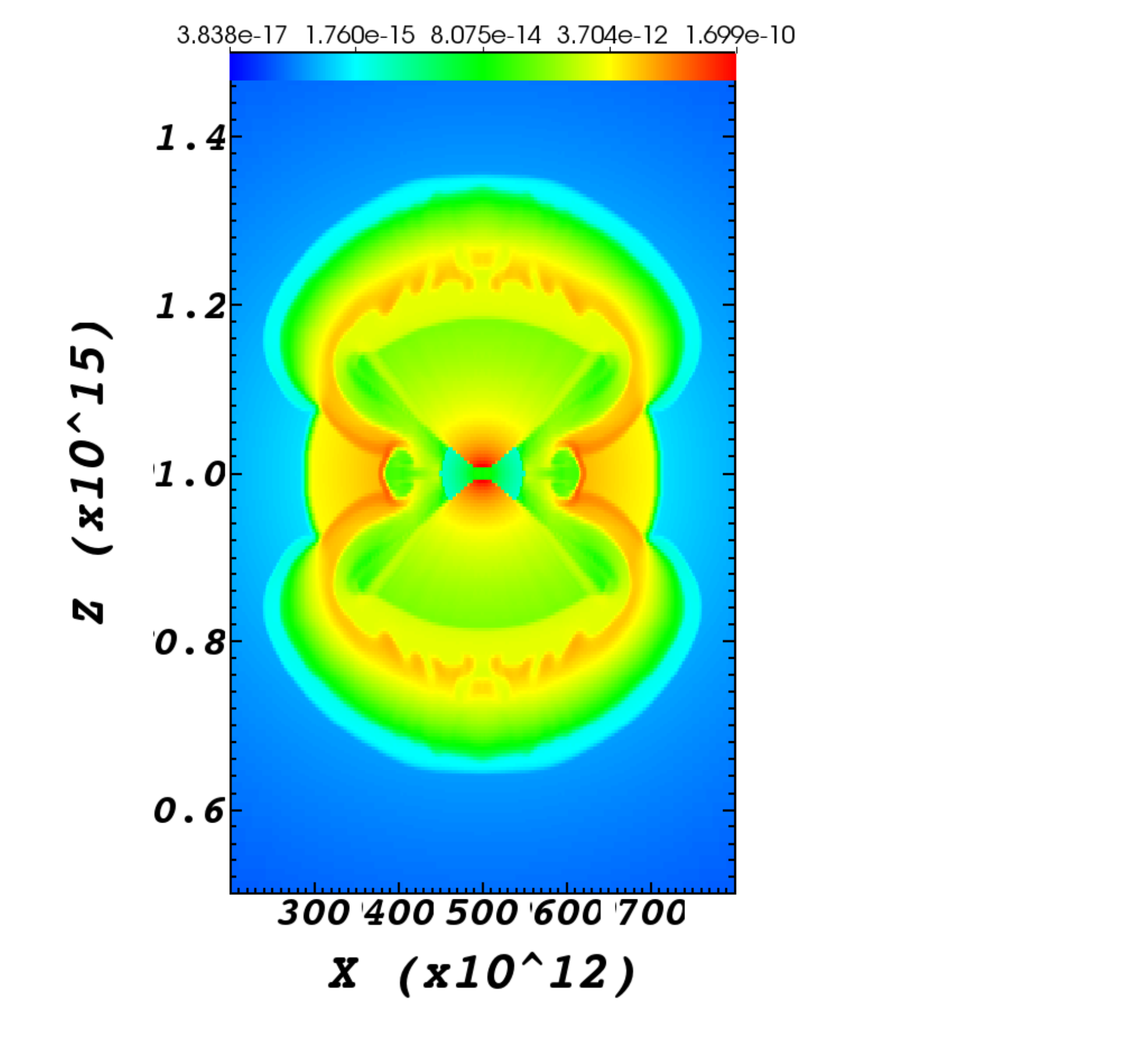}}
\centering
\caption{The density maps at $t=57.6 \days$ for three different runs with three values of the adiabatic index:
(a) $\gamma=1.02$; (b) $\gamma=1.05$; (c) $\gamma=5/3$, in logarithmic scale.
Color coding is in units of $\g \cm^{-3}$, and units on the axes are in $\cm$. }
\label{fig:compare_gama}
\end{figure}

We mimic the cooling via photon diffusion by  taking the adiabatic index to be $\gamma<5/3$.
To explore the role of different cooling to flow time ratio we also study cases with values of $\gamma = 1.02$, $1.05$, and $5/3$.
The density maps at $t= 57.6 \yr$ for these cases are presented in Fig. \ref{fig:compare_gama}.
As expected, when radiative cooling is efficient the nebula has a lower velocity, and the instabilities are less developed.
For $\gamma=1.05$ instability fingers can still be identified, but barely so for the run with $\gamma=1.02$.
These instabilities might be the source for the clumpy structure observed in NGC~6302 and similar PNe.

\section{SUMMARY}
\label{sec:summary}

We perform 3D hydrodynamical simulations of a very-short duration jet-shell interaction.
Such an interaction  can take place when for some reason the AGB star goes through an unstable phase and eject within a period of months-years
a large amount of mass $\sim 0.01-1 M_\odot$ \citep{SokerKashi2012}. This ejection is most likely to be aided by a close companion.
In our scenario the companion accretes a large fraction of the ejected mass and launches two opposite jets, in one or more episodes \citep{SokerKashi2012}.
During the planetary nebula (PN) and pre-PN phases a structure of two lobes having a linear distance-velocity relation will be observed.

In section \ref{sec:diffusion} we find that in the jet-shell interaction region the photon-diffusion time is not
much shorter, and even longer, than the gas expansion time.
This implies that the radiative cooling process is relatively slow, and the shocked gas can stay hot for the duration of the acceleration time
of the shell by the jets.
This regime of optically-thick interaction is not traditionally considered for PN formation.
We make importance of this new regime, and simulated the interaction in a simple manner, where we mimic
the slow radiative cooling process by taking the adiabatic index to be $\gamma<5/3$.
The results are presented in Figs. \ref{fig:General} - \ref{fig:compare_gama}.

We simulate interactions with physical parameters such that the dense shell is denser than the post-shock jets' gas.
Such a flow is expected to be Rayleigh-Taylor unstable.
Our main finding indeed, is that as the jets interact with the dense shell instabilities develop and a structure
containing dense `fingers' is formed, as is expected in Rayleigh-Taylor or thin-shell instabilities.
At much later times a PN with clumpy lobes that have a linear distance-velocity relation will be observed.
This process might account for the formation of bipolar PNe with clumpy lobes, such as NGC 6302.
To better compare model with observations we are planing to extended our study to include late phases of the flow,
and to expand the parameters space of the physical variables.

The energy radiated during the months to years duration of such an event will appear as an
intermediate-luminosity optical transient ({ILOT}; also termed Red Nova) event.
In our scenario there are two sources for the radiated energy of the  {ILOT}: the accretion onto the companion and the collision of the jets with the shell.
\cite{Ivanova2013} consider the entrance of a system to a common envelope phase, and argue that most of the energy might come
from the recombination energy of the ejected gas. In our scenario (and in most  {ILOTs}) the recombination energy of the ejected envelope is negligible,
and the companion might or might not enter the envelope of the AGB star.

We thank Bruce Balick and Amit Kashi for helpful comments.
This research was supported by the Asher Fund for Space Research
at the Technion, and the US-Israel Binational Science Foundation.
We thank an anonymous referee for helpful comments.



\footnotesize

\begin{thebibliography}

\bibitem[Akashi et al.(2008)]{Akashietal2008} Akashi, M., Meiron, Y., \& Soker, N.\ 2008, \na, 13, 563

\bibitem[Akashi \& Soker(2008)]{AkashiSoker2008} Akashi, M., \& Soker, N.\ 2008, \mnras, 391, 1063

\bibitem[Alcolea et al.(2001)]{Alcolea2001} Alcolea, J., Bujarrabal, V., S{\'a}nchez Contreras, C., Neri, R., \& Zweigle, J.\ 2001, \aap, 373, 932

\bibitem[Arnett(1979)]{Arnett1979} Arnett, W.~D.\ 1979, \apjl, 230, L37

\bibitem[Balick et al.(2013)]{Balicketal2013} Balick, B. Huarte-Espinosa,
M., Frank, A., Gomez, T., Alcolea, J., Corradi, R.L.M., \& Vinkovic, D.\
2013, (arXiv:1305.5304) 

\bibitem[Barbary et al.(2009)]{Barbary2009} Barbary, K., et al. 2009, \apj, 690, 1358

\bibitem[Berger et al.(2009)]{Berger2009} Berger, E., et al. 2009, \apj, 699, 1850

\bibitem[Berger et al.(2011)]{Berger2011}Berger, E., Foley, R., \& Soderberg, A. 2011, The Astronomer�s Telegram, 3467

\bibitem[Bond et al.(2009)]{Bond2009} Bond, H.~E., Bedin, L.~R., Bonanos, A.~Z., Humphreys, R.~M., Monard, L.~A.~G.~B., Prieto, J.~L., \& Walter, F.~M.\ 2009, \apjl, 695, L154

\bibitem[Botticella et al.(2009)]{Botticella2009} Botticella, M.~T., et al. 2009, \mnras, 398, 1041

\bibitem[Boumis \& Meaburn(2013)]{BoumisMeaburn2013} Boumis, P., \& Meaburn, J.\ 2013, \mnras, 430, 3397

\bibitem[Bujarrabal et al.(1998)]{Bujarrabal1998} Bujarrabal, V., Alcolea, J., Sahai, R., Zamorano, J., \& Zijlstra, A.~A.\ 1998, \aap, 331, 361

\bibitem[Bujarrabal et al.(2002)]{Bujarrabal2002} Bujarrabal, V., Alcolea, J., S{\'a}nchez Contreras, C., \& Sahai, R.\ 2002, \aap, 389, 271

\bibitem[Dennis et al.(2008)]{Dennisetal2008} Dennis, T.~J., Cunningham, A.~J., Frank, A., Balick, B., Blackman, E.~G., \& Mitran, S.\ 2008, \apj, 679, 1327

\bibitem[Dennis et al.(2009)]{Dennisetal2009} Dennis, T.~J., Frank, A., Blackman, E.~G., De Marco, O., Balick, B., \& Mitran, S.\ 2009, \apj, 707, 1485

\bibitem[Fryxell et al.(2000)]{Fryxell2000} Fryxell, B., Olson, K., Ricker, P., et al.\ 2000, \apjs, 131, 273

\bibitem[Gaetz et al.(1988)]{Gaetz88} Gaetz, T.~J., Edgar, R.~J., \& Chevalier, R.~A.\ 1988, \apj, 329, 927

\bibitem[Garc{\'{\i}}a-Arredondo \& Frank(2004)]{GarciaFrank2004} Garc{\'{\i}}a-Arredondo, F., \& Frank, A.\ 2004, \apj, 600, 992

\bibitem[Guerrero et al.(2004)]{Guerreroetal2004} Guerrero, M.~A., Chu, Y.-H., \& Miranda, L.~F.\ 2004, \aj, 128, 1694

\bibitem[Hajduk et al.(2013)]{Hajduketal2013} Hajduk, M., van Hoof, P.~A.~M., \& Zijlstra, A.~A.\ 2013, \mnras, 1133

\bibitem[Huarte-Espinosa et al.(2012)]{Huarte-Espinosa2012} Huarte-Espinosa, M., Frank, A., Balick, B., Blackman, E. G., De Marco, O.,
 Kastner, J. H., \& Sahai, R.\ 2012, \mnras, 424, 2055

\bibitem[Humphreys et al.(2011)]{Humphreysetal2011} Humphreys, R.~M., Bond, H.~E., Bedin, L.~R., Bonanos, A. Z.,
    Davidson, K., Berto Monard, L. A. G., Prieto, J. L., \& Walter, F. M.\ 2011, \apj, 743, 118

\bibitem[Ivanova et al.(2013)]{Ivanova2013} Ivanova, N., Justham, S., Avendano Nandez, J.~L., \& Lombardi, J.~C.\ 2013, Science, 339, 433

\bibitem[Kashi et al.(2010)]{Kashi2010} Kashi, A., Frankowski, A., \& Soker, N.\ 2010, \apjl, 709, L11

\bibitem[Kashi \& Soker(2010a)]{KashiSoker2010a} Kashi, A., \& Soker, N.\ 2010a, \apj, 723, 602

\bibitem[Kashi \& Soker(2010b)]{KashiSoker2010b} Kashi, A., \& Soker, N.\ 2010b, (arXiv:1011.1222)

\bibitem[Kasliwal et al.(2011)]{Kasliwal2011} Kasliwal, M.~M., et al. 2011, \apj, 730, 134

\bibitem[Kastner et al.(1992)]{Kastner1992} Kastner, J.~H., Weintraub, D.~A., Zuckerman, B., Becklin, E.~E., McLean, I., \& Gatley, I.\ 1992, \apj, 398, 552

\bibitem[Kastner et al.(1998)]{Kastner1998} Kastner, J.~H., Weintraub, D.~A., Merrill, K.~M., \& Gatley, I.\ 1998, \aj, 116, 1412

\bibitem[Kochanek(2011)]{Kochanek2011} Kochanek, C.~S.\ 2011, \apj, 741, 37

\bibitem[Kulkarni \& Kasliwal(2009)]{KulkarniKasliwal2009} Kulkarni, S.~R., \& Kasliwal, M.~M.\ 2009, astro2010: The Astronomy and Astrophysics Decadal Survey, 2010, 165

\bibitem[Lee et al.(2009)]{Leeetal2009} Lee, C.-F., Hsu, M.-C., \& Sahai, R.\ 2009, \apj, 696, 1630

\bibitem[Lee \& Sahai(2003)]{LeeSahai2003} Lee, C.-F., \& Sahai, R.\ 2003, \apj, 586, 319

\bibitem[Lee \& Sahai(2004)]{LeeSahai2004} Lee, C.-F., \& Sahai, R.\ 2004, \apj, 606, 483

\bibitem[Levesque et al.(2013)]{Levesque2013} Levesque, E.~M., Stringfellow, G.~S., Ginsburg, A.~G., Bally, J., \& Keeney, B.~A.\ 2013, arXiv:1211.4577

\bibitem[Mason et al.(2010)]{Mason2010} Mason, E., Diaz, M., Williams, R.~E., Preston, G., \& Bensby, T.\ 2010, \aap, 516, A108

\bibitem[Matsuura et al.(2005)]{Matsuura2005} Matsuura, M., Zijlstra, A.~A., Molster, F.~J., Waters, L.~B.~F.~M., Nomura, H., Sahai, R., \& Hoare, M.~G.\ 2005, \mnras, 359, 383

\bibitem[Mauerhan et al.(2013)]{Mauerhan2013} Mauerhan, J.~C., et al.\ 2013, \mnras, 430, 1801

\bibitem[McLeod \& Whitworth(2013)]{McLeod2013} McLeod,A. D. \& Whitworth. A. P.\ 2013,  \mnras, 431, 710

\bibitem[Meaburn et al.(2008)]{Meaburn2008} Meaburn, J., Lloyd, M., Vaytet, N.~M.~H., \& L{\'o}pez, J.~A.\ 2008, \mnras, 385, 269

\bibitem[Mould et al.(1990)]{Mould1990} Mould, J., et al. 1990, \apjl, 353, L35

\bibitem[Ofek et al.(2008)]{Ofek2008} Ofek, E.~O., et al. 2008, \apj, 674, 447

\bibitem[Pastorello et al.(2010)]{Pastorello2010} Pastorello, A., et al. 2010, \mnras, 408, 181

\bibitem[Prieto et al.(2009)]{Prieto2009} Prieto, J.~L., Sellgren, K., Thompson, T.~A., \& Kochanek, C.~S.\ 2009, \apj, 705, 1425

\bibitem[Rau et al.(2007)]{Rau2007} Rau, A., Kulkarni, S.~R., Ofek, E.~O., \& Yan, L.\ 2007, \apj, 659, 1536

\bibitem[Sahai et al.(2006)]{Sahai2006} Sahai, R., Young, K., Patel, N.~A., S{\'a}nchez Contreras, C., \& Morris, M.\ 2006, \apj, 653, 1241

\bibitem[Sahai et al.(2003)]{Sahai2003} Sahai, R., Zijlstra, A., S{\'a}nchez Contreras, C., \& Morris, M.\ 2003, \apjl, 586, L81

\bibitem[S{\'a}nchez Contreras et al.(2004)]{Contreras2004} S{\'a}nchez Contreras, C., Gil de Paz, A., \& Sahai, R.\ 2004, \apj, 616, 519

\bibitem[Smith et al.(2009)]{Smith2009} Smith, N., et al. 2009, \apjl, 697, L49

\bibitem[Smith et al.(2011)]{Smithetal2011} Smith, N., Li, W., Silverman, J.~M., Ganeshalingam, M., \& Filippenko, A.~V.\ 2011, \mnras, 415, 773

\bibitem[Smith et al.(2010)]{Smithetal2010} Smith, N., et al.\ 2010, \aj, 139, 1451

\bibitem[Soker \& Kashi(2011)]{SokerKashi2011} Soker, N., \& Kashi, A.\ 2011 (arXiv:1107.3454)

\bibitem[Soker \& Kashi(2012)]{SokerKashi2012} Soker, N., \& Kashi, A.\ 2012, \apj, 746, 100

\bibitem[Soker \& Kashi(2013)]{SokerKashi2013} Soker, N., \& Kashi, A.\ 2013, \apjl, 764, L6

\bibitem[Szyszka et al.(2011)]{Szyszka2011} Szyszka, C., Zijlstra, A.~A., \& Walsh, J.~R.\ 2011, (arXiv:1105.3381)

\bibitem[Trammell \& Goodrich(1996)]{Trammell1996} Trammell, S.~R., \& Goodrich, R.~W.\ 1996, \apjl, 468, L107

\bibitem[Tylenda et al.(2013)]{Tylendaetal2013} Tylenda, R., Kaminski, T., Udalski, A., et al.\ 2013, arXiv:1304.1694

\bibitem[Ueta et al.(2007)]{Ueta2007} Ueta, T., Murakawa, K., \& Meixner, M.\ 2007, \aj, 133, 1345

\bibitem[Wright et al.(2011)]{Wright2011} Wright, N.~J., Barlow, M.~J., Ercolano, B., \& Rauch, T.\ \mnras, 2011, (arXiv:1107.4554)


\end{thebibliography}
\end{document}